\documentclass[twocolumn,showpacs,preprintnumbers,amsmath,amssymb]{revtex4}

\usepackage{graphicx}
\usepackage{dcolumn}
\usepackage{bm}

\begin{document}

\preprint{APS/123-QED}

\title{On different cascade-speeds for longitudinal and transverse velocity increments}

\author{M. Siefert}
\author{J. Peinke}%
 \email{peinke@uni-oldenburg.de}
 \homepage{http://www.uni-oldenburg.de/hydro}
\affiliation{%
Institut f\"ur Physik, Carl-von-Ossietzky Universit\"at Oldenburg, D-26111 Oldenburg, Germany 
}%


\date{\today}

\begin{abstract}
We address the problem of differences between longitudinal and transverse velocity increments in isotropic small scale turbulence. The relationship of these two quantities is analyzed experimentally by means of stochastic Markovian processes leading to a phenomenological Fokker- Planck equation from which a generalization of the K\'arm\'an equation is derived. From these results, a simple relationship between longitudinal and transverse structure functions is found which explains the difference in the scaling properties of these two structure functions.

\end{abstract}

\maketitle

Substantial details of the complex statistical behaviour of fully
developed turbulent flows are still unknown, cf. \cite{Monin,Frisch,Tsinober,nelkin00}.
 One important task is to understand intermittency, i.e. finding unexpected frequent occurences of large fluctuations of the local velocity on small length scales.
 In the last years, the differences of velocity fluctuations in different spatial directions have attracted considerable attention as a main issue of the problem of small scale turbulence, see for example \cite{antonia00a,arad98,chen97a,dhruva97,gotoh02,grossmann97a,laval01,water99,hill01a}.
For local isotropic turbulence, the statistics of velocity increments $\left[{\bf v}({\bf x}+{\bf r})-{\bf v}({\bf x})\right]{\bf e}$  as a function of the length scale $r$ is of interest. 
Here, ${\bf e}$ denotes a unit vector. We denote  with $u(r)$ the longitudinal increments (${\bf e}$ is parallel to ${\bf r}$) and with $v(r)$ transverse increments (${\bf e}$ is orthogonal to ${\bf r}$)
\footnote{Note we focus on transverse increments for which ${\bf r}$ is in direction to the mean flow and $v$ is perpendicular to the mean flow. An other possibility is to chose ${\bf r}$ perpendicular to the mean flow, which possibly results in a different behavior in case of anisotropy \cite{shen02,water99}.}.

In a first step, this statistics is commonly investigated by means of its
moments $\langle u^n(r) \rangle$ or $\langle v^n(r) \rangle$, the so-called  velocity
structure functions.  Different theories and models try to explain the shape of the structure functions cf. \cite{Frisch}.
Most of the works examine the scaling of the structure function, $\langle u^n\rangle \propto r^{\xi_l^n}$, and try to explain intermittency, expressed by $\xi_l^n-n/3$ the deviation from Kolmogorov theory of 1941 \cite{kolmogorov41,kolmogorov62}. For the corresponding transverse quantity we write  $\langle v^n\rangle \propto r^{\xi_t^n}$.
There is strong evidence that there are fundamental differences in the statistics of the longitudinal increments $u(r)$ and transverse increments $v(r)$. Whereas there were some contradictions initially, there is evidence now that the transverse scaling shows stronger intermittency even for high Reynolds numbers \cite{shen02,gotoh02}.
 
A basic equation which relates both quantities is derived by K\'arm\'an and Howarth \cite{karman38}. Assuming incompressibilty and isotropy, the so called {\em first K\'arm\'an equation} is obtained:
\begin{eqnarray}\label{karman1}
-r\frac{\partial}{\partial r}\langle u^2\rangle=2\langle u^2\rangle-2\langle v^2\rangle .
\end{eqnarray}
Relations between structure functions become more and more complicated with higher order, including also pressure terms \cite{hill01a,hill01c}.

In this paper, we focus on a different approach to characterize spatial multipoint correlations via multi-scale statistics. Recently it has been shown that it is possible to get access to the joint probability distribution $p(u(r_1),u(r_2),\dots,u(r_n))$ via a Fokker-Planck equation, which can be estimated directly from measured data \cite{friedrich97a,friedrich97b,marcq01}. For a detailed presentation see  \cite{renner01a}. This method is definitely more general than the above mentioned analysis by structure functions, which characterize only the simple scale statistics $p(u(r))$ or $p(v(r))$. The Fokker-Planck method has attracted interest and was applied to different problems of the complexity of turbulence like energy dissipation \cite{naert97,renner03,marcq98}, universality turbulence \cite{renner02} and others \cite{davoudi99,frank03,hosokawa02,laval01,ragwitz01,schmitt01}.  The Fokker-Planck equation (here written for vector quantities) reads as
\begin{eqnarray}\label{fpgl}
\lefteqn{-r\frac{\partial}{\partial r}p({\bf u},r|{\bf u_0},r_0)=}\\\nonumber
&&\left(-\sum^n_{i=1}\frac{\partial}{\partial u_i}D^{(1)}_i +\sum_{i,j=1}^n\frac{\partial^2}{\partial u_i \partial u_j}D^{(2)}_{ij}\right)p({\bf u},r|{\bf u_0},r_0)\nonumber.
\end{eqnarray}
($i$ denotes the spatial component of $\bf u$, we fix
 $i=1$ for the longitudinal and $i=2$ for the transverse increments.)
This representation of a stochastic process is different from the usual one: instead of the time $t$, the independent variable is the scale variable $r$. The minus sign appears from the development of the probability distribution from large to small scales.  In this sense, this Fokker- Planck equation may be considered as an equation for the dynamics of the cascade, which describes how the increments evolve from large to small scales under the influence of deterministic ($D^{(1)}$) and noisy ($D^{(2)}$) forces.
 The whole equation is multiplied without loss of generality by $r$ to get power laws for the moments in a more simple way, see also Eq. (\ref{karman1}). 
Both coefficients, the so-called drift term $D^{(1)}_i({\bf u},r)$ and  diffusion term $D^{(2)}_{ij}({\bf u},r)$, can be estimated directly from the measured data using its mathematical definition, see Kolmogorov 1931 \cite{kolmogorov31} and \cite{Risken,renner01a,ragwitz01,friedrich02}. With the notation $\Delta U_i(r,\Delta r):=U_i(r-\Delta r) - u_i(r)$ the definitions read as:
\begin{equation}\label{d1}
D^{(1)}_i({\bf u},r)=
\lim_{\Delta r \to 0}\frac{r}{\Delta r}\langle \Delta U_i(r,\Delta r)\rangle|_{{\bf U}(r)={\bf u}(r)}\nonumber,
\end{equation}
\begin{eqnarray}\label{d2}
\lefteqn{D^{(2)}_{ij}({\bf u},r)=}\\
&&\lim_{\Delta r \to 0}\frac{r}{2\Delta r}\langle \Delta U_i(r,\Delta r) \Delta U_j(r,\Delta r)\rangle|_{{\bf U}(r)={\bf u}(r)}\nonumber.
\end{eqnarray}

Here we extend the analysis to a two dimensional Markov process, relating the longitudinal and transverse velocity increments to each other. The resulting Fokker-Planck equation describes the joint probability distribution $p(u({r_1}),v(r_1);\dots;u(r_n),v(r_n))$. Knowing the Fokker-Planck equation, hierarchical equations for any structure function $\langle f(u,v)\rangle=\int  f(u,v)p(u,v,r)dudv$ can be derived by integrating Eq. (\ref{fpgl}):
\begin{eqnarray}\label{StrukturGl}
\lefteqn{-r\frac{\partial}{\partial r}\langle u^mv^n\rangle =}\\
&+&m \langle u^{m-1}v^n D^{(1)}_1(u,v,r) \rangle+n\langle u^mv^{n-1}D^{(1)}_2(u,v,r) \rangle\nonumber\\&+&\frac{m(m-1)}{2}\langle u^{m-2}v^n D^{(2)}_{11}(u,v,r)\rangle\nonumber\\&+&\frac{n(n-1)}{2}\langle u^{m}v^{n-2} D^{(2)}_{22}(u,v,r)\rangle\nonumber\\ &+&mn\langle u^{m-1}v^{n-1}D^{(2)}_{12}(u,v,r)\rangle,\nonumber
\end{eqnarray}
which we take as a generalization of the K\'arm\'an equation.

Next, we describe the experiment used for the subsequent analysis. The data set consists of $1.25\cdot10^8$ samples of the local velocity measured in the wake behind a cylinder (cross section: $D$=20mm) at a Reynolds' number of 13236 and a Taylor- based Reynolds' number of 180. The measurement was done with 
a X-hotwire placed 60 cylinder diameters behind the cylinder. The component $u$ is measured along the mean flow direction,
the component $v$ transverse to the mean flow  is orthogonal to the cylinder axis. We use Taylor's hypothesis of frozen turbulence to convert time lags into spatial displacements.  With the sampling
frequency of 25kHz and a mean velocity of 9.1 m/s, the spatial
resolution doesn't resolve the Kolmogorov length but the Taylor length $\lambda=4.85$ mm. The integral length is $L=$137 mm.

From these experimental data the drift and diffusion coefficients are estimated according to eqs. (\ref{d1})  as described in \cite{friedrich02}, see also \cite{ETC9}. As an example, the diffusion coefficient $D^{(2)}_{11}(u,v,r=L/4)$ is shown in Fig. \ref{fig:D2}.
 To use the results in an analytical way, the drift and diffusion coefficient can be well approximated by the following low dimensional polynoms, which will be verified by reconstructed structure functions as it is shown below: 
\begin{eqnarray}\label{D_Polynom}
D^{(1)}_1(u,v,r)&=&d^u_1(r)u\\\nonumber
D^{(1)}_2(u,v,r)&=&d^v_2(r)v\\\nonumber
D^{(2)}_{11}(u,v,r)&=&d_{11}(r)+d^u_{11}(r)u+d^{uu}_{11}(r)u^2+d^{vv}_{11}(r)v^2\\\nonumber
D^{(2)}_{22}(u,v,r)&=&d_{22}(r)+d^u_{22}(r)u+d^{uu}_{22}(r)u^2+d^{vv}_{22}(r)v^2\\\nonumber
D^{(2)}_{12}(u,v,r)&=&d_{12}(r)+d^u_{12}(r)u+d^{uv}_{12}(r)uv . \nonumber
\end{eqnarray}
In order to show that the Fokker-Planck equation with these drift and diffusion coefficients can well characterize the increment's statistics, one has to verify  that the evolution process of $u(r$) and $v(r)$ is a Markov process and that white and Gaussian distributed noise is involved.  The Markov property can be tested directly via its definition by using conditional probability densities \cite{renner01a} or by looking at the correlation of the noise of the Langevin equation  \cite{marcq01}.  For our case we have verified that the one- dimensional processes of the longitudinal and transverse increments are Markovian \footnote{ to be published }, thus the two- dimensional processes should be Markovian, too {\cite{Risken}. As an alternative approach to verify the validity of the Fokker-Planck equation we have solved numerically the hierarchical eq. (\ref{StrukturGl}) for $\langle u^m \rangle$ using the above mentioned coefficients. To close the equation, we have used the moment $\langle u^{m-2}v^2\rangle$ from the experimental data. In Fig. \ref{fig:verific} the integrated longitudinal structure functions  are given in comparison with the structure functions directly calculated from the data (with $n=$2, 4, 6 and $m=$0). These results we take as  the evidence  that the Fokker-Planck equation characterizes the data well and can be used for further interpretations.

\begin{figure}
\includegraphics[width=3.0in]{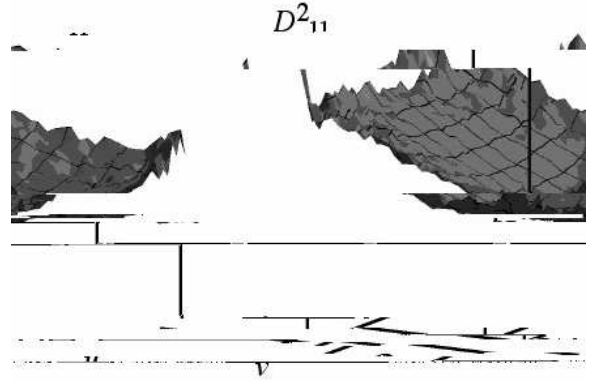}
\caption{\label{fig:D2} Diffusion coefficient $D^{(2)}_{11}(u,v,r)$ for $r=L/4$ estimated from experimental data. Note that the quadratic contributions are responsible for intermittency. The asymmetry in $u$-direction is related to the non-vanishing skewness of the probability distribution of longitudinal velocity increments.}
\end{figure}

\begin{figure}
\includegraphics[width=2.5in]{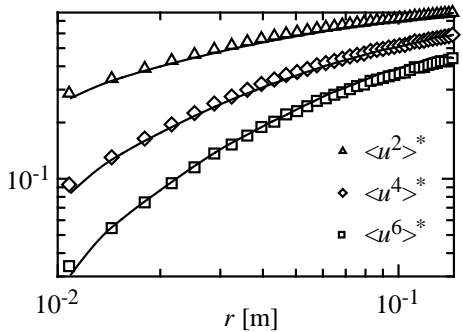}
\caption{\label{fig:verific}Even longitudinal structure functions up to order 6 calculated directly from data (symbols) are compared to the results obtained from numerical solutions of Eq. (\ref{StrukturGl}), using the experimentally estimated coefficients of the Fokker-Planck equation (solid lines; see also Eq. (\ref{D_Polynom}) and Fig. \ref{fig:D2}).}
\end{figure}

For the drift coefficient, which is the deterministic part of the cascade dynamics, the process decouples for the different directions.
The drift and diffusion coefficients are symmetric with respect to $v\to -v$. Remark, in contrast to the statistics of the longitudinal increments, the transverse one is symmetric and show for example no skewness $\langle v^3\rangle=0$.
Furthermore, quadratic terms occur in the diffusion coefficients. Intermittency results from the quadratic terms $d_{11}^{uu}$ and $d_{22}^{vv}$, all other terms act against intermittency.

The $r$-dependence of related longitudinal and transverse $d$-coefficients ($d_1^u$ and $d_2^v$ etc.) coincides if the abscissa are rescaled:  $d_\textrm{long}(r)\approx d_\textrm{transv}(\frac{2}{3}r)$, see Fig. \ref{fig:d_r}. The only exception is the coefficient $d_{22}^{vv}(r)$, whereas $d_{11}^{uu}(r) \approx d_{11}^{vv}(r) \approx d_{22}^{uu}(r) \approx const. $.
We interpret this phenomenon as a faster cascade for the transverse increments.
 It can be seen from the hierarchical equation (\ref{StrukturGl}) that this property goes over into structure functions of arbitrary even order, $\langle v^n(r)\rangle \approx \langle u^n(\frac{3}{2}r)\rangle$.  Only the small coefficients $d_{11}^u$ and $d_{22}^u$ break this symmetry, because they belong to different odd, and therefore small, moments.
In Fig. \ref{fig:structfct}, the structure functions of order 2, 4 and 6 are plotted with respect to this rescaled length $r$. The structure functions are normalized by $\langle \alpha^n\rangle^*=\langle \alpha^n\rangle\frac{(n/2)!}{n!u^n_\textrm{rms}}$, with $\alpha$ either $u$ or $v$.

\begin{figure}
\includegraphics[width=3.5in]{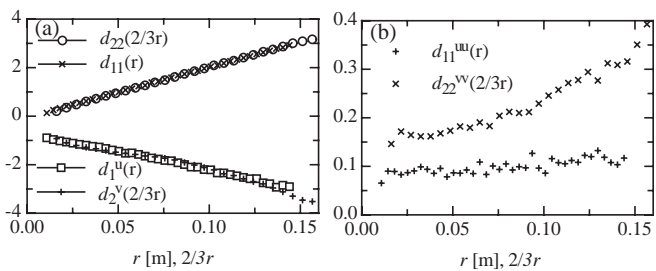}
\caption{\label{fig:d_r} The expansion coefficients of the drift and diffusion coefficients in dependence of the scale $r$. The abscissa are rescaled for the transverse coefficients. The corresponding longitudinal and transverse coefficients coincide which each other a) apart from the intermittency terms b).}  
\end{figure}

\begin{figure}
\includegraphics[width=2.5in]{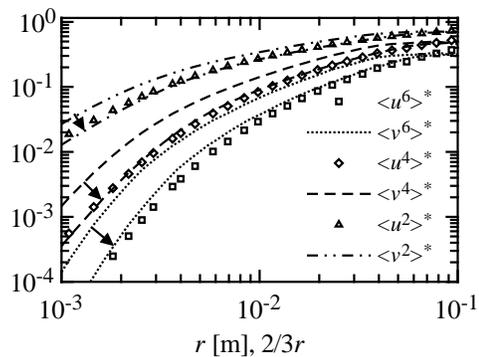}
\caption{\label{fig:structfct} Longitudinal (symbols) and  corresponding transverse (lines) structure functions. The arrows point from the structure function $\langle v^n(r) \rangle$ to the corresponding structure function $\langle v^n(2/3\;r) \rangle$ with contracted abscissa. The transverse structure functions with a contracted abscissa are close to the longitudinal ones.}
\end{figure}

The observed phenomena are consistent with the K\'arm\'an equation (\ref{karman1}), if the K\'arm\'an equation is interpreted as a Taylor expansion
\begin{eqnarray}
\langle v^2(r)\rangle&=&\langle u^2(r)\rangle+\frac{1}{2}r\frac{\partial}{\partial r}\langle u^2(r)\rangle\\&\approx&\langle u^2(r+\frac{1}{2}r)\rangle=\langle u^2(\frac{3}{2}r)\rangle.
\end{eqnarray}

Next, let us suppose that the structure functions scale with a power law, $\langle v^n(r)\rangle=c_t^nr^{\xi_t^n}$ and $\langle u^n(r)\rangle=c_l^nr^{\xi_l^n}$, even though our measured structure functions are still far away from showing an ideal  scaling behavior 
\cite{renner02}. With exemption of the differences between $d_{11}^{uu}(r)$ and $d_{22}^{vv}(r)$, we can relate the structure functions according to the above mentioned rescaling: $\langle v^n(r)\rangle=\langle u^n(\frac{3}{2}r)\rangle=c_t^nr^{\xi_t^n}=c_l^n(\frac{3}{2}r)^{\xi_l^n}$. We end up with the relation $\xi_l^n=\xi_t^n$ and $\frac{c_t^n}{c_l^n}=\left(\frac{3}{2}\right)^{\xi_l^n}$.  Note that the $c^n$ constants are related to the Kolmogorov constants.  For $n=2$ and $4$ we obtain $c_t^2/c_l^2\approx 1.33$ and $c_t^4/c_l^4\approx 1.72$, which deviates less than  3\% from the value of $c_t^2/c_l^2 = 4/3$ and  $c_t^4/c_l^4=16/9$ given in \cite{antonia97c}. 

At last we discuss the use of ESS (extended self-similarity \cite{benzi93b} ) with respect to transverse velocity components. In \cite{pearson01,dhruva97,antonia97a} the authors plot  $\langle v^n\rangle$ against $\langle |u^3|\rangle$ and obtain that the transverse exponents is smaller than the longitudinal one, $\xi^n_{t}< \xi^n_{l}$. In Fig. \ref{fig:ess} the fourth structure functions are plotted against $\langle |u^3|\rangle$, which shows clearly that $\xi^4_{t} < \xi^4_{l}$. If the transverse structure function is plotted as a function of $\langle |u^3(\frac{3}{2}r)| \rangle $, this discrepancy vanishes. Notice that these properties are due to a none existing scaling behavior. It is evident that our rescaling does not change the exponents  in case of pure scaling behavior.

\begin{figure}
\includegraphics[width=2.5in]{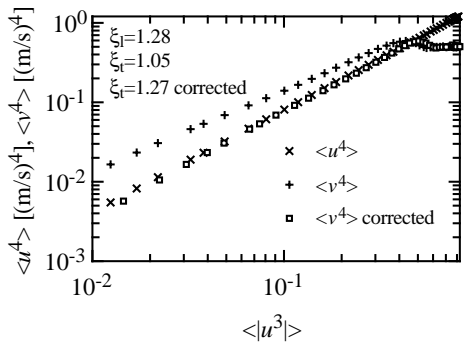}
\caption{\label{fig:ess} The fourth order longitudinal (x) and transverse (+) structure function is plotted against $\langle |u|^3\rangle$, the third moment of the {\em longitudinal} increments' modulus (ESS plot). If the abscissa of the transverse structure function is rescaled by a factor of $2/3$, both curves fall one upon the other (squares). The differences between the exponents also vanish.}
\end{figure}

To conclude the paper, we have presented experimental evidence
 that the statistics of longitudinal and transverse increments is dominated by a difference in the "speed of cascade"  expressed by its r dependence. Rescaling the r dependence of the transverse increments by a factor $\frac{2}{3}$ fades the main differences away. Thus the longitudinal and transverse structure functions up to order 6 coincide well. A closer look at the coefficients of the stochastic process estimated from our data shows that the multiplicative noise term for the transverse increments $d_{22}^{vv}$ and the symmetry breaking terms $d_{11}^u$ and $d_{22}^u$ do not follow this rescaling. These coefficients may be a source of differences for the two directions, but for our data analyzed here this effect is still very small.

Finally, we could show that our findings on the rescaling are consistent with the  K\'arm\'an equation and that  longitudinal and transverse Kolmogorov constants of the structure functions up to order four can be related consistently with our results.

We acknowledge fruitful discussions with R. Friedrich and teamwork with M. Karth. This work was supported by the DFG-grant Pe 478/9.

\bibliography{lit}
\end{document}